\documentclass{article}

\usepackage[margin=1.3in]{geometry}
\usepackage[utf8]{inputenc}
\usepackage{amsmath}
\usepackage{amssymb}
\usepackage{amsthm}
\usepackage{graphicx}
\usepackage{subcaption}
\usepackage{array,multirow}

\usepackage{setspace}

\usepackage[style=apa,backend=biber]{biblatex} % Imports biblatex package
\usepackage{caption}
\usepackage{subcaption}
\usepackage{float}
\usepackage{authblk}

\author[1]{David Swanson}
\author[2]{Alexander Sherry}
\author[2]{Chad Tang}
\affil[1]{Department of Biostatistics,}
\affil[2]{Department of Radiation Oncology, \hspace{-4cm}  \newline University of Texas MD Anderson Cancer Center}

\title{Variance component mixture modelling for longitudinal T-cell receptor clonal dynamics}
\date{February 2025}

\begin{document}

\maketitle

\abstract{
Studies of T cells and their clonally unique receptors have shown promise in elucidating the association between immune response and human disease. Methods to identify T-cell receptor clones which expand or contract in response to certain therapeutic strategies have so far been limited to longitudinal pairwise comparisons of clone frequency with multiplicity adjustment.  Here we develop a more general mixture model approach for arbitrary follow-up and missingness which partitions dynamic longitudinal clone frequency behavior from static. While it is common to mix on the location or scale parameter of a family of distributions, the model instead mixes on the parameterization itself, the dynamic component allowing for a variable, Gamma-distributed Poisson mean parameter over longitudinal follow-up, while the static component mean is time invariant. Leveraging conjugacy, one can integrate out the mean parameter for the dynamic and static components to yield distinct posterior predictive distributions whose expressions are a product of negative binomials and a single negative multinomial, respectively, each modified according to an offset for receptor read count normalization.
% and interpretation as a clone frequency. 
% the empirical Bayes h
An EM-algorithm is developed to estimate hyperparameters and component membership, and validity of the approach is demonstrated in simulation. The model identifies a statistically significant and clinically relevant increase in TCR clonal dynamism among metastasis-directed radiation therapy in a cohort of prostate cancer patients.
% \vspace{0.4cm}

\noindent
{\bf keywords:} Bayesian conjugacy, hierarchical model, mixture model, EM algorithm, t-cell receptor}

\section{Introduction}

% talk about missingness advantage

% TEST
% The value of my variable is \myvariable.
% TEST

Immune response plays a central role in human health and addressing the many foreign agents people are regularly exposed to.  T-cells are an important piece of one's adaptive immune system, in part by recognizing antigens with receptors that exhibit an exceptional degree of diversity via highly variable gene recombination in portions of the receptor \parencite{hey_analysis_2023,chiou_global_2021}. Study of T-cell receptor (TCR) repertoires has gained attention in recent years because immunosequencing receptors allows one to examine TCR clonal dynamics and their association with biomarkers, disease, and other aspects of human biology \parencite{teng_analysis_2022, rosati_overview_2017, luo_characteristics_2022}.  
% \citeauthor{dempster_maximum_1977}.  
Clones of receptors binding an antigen tend to expand, which can be measured by calculating the proportion of a certain clone within the repertoire using sequencing technologies and examining evolution of the quantity over time \parencite{chen_decreased_2021}.  Other clones contract or are relatively static proportionally within the repertoire, and one can identify~ these distinct dynamics longitudinally \parencite{dewitt_dynamics_2015}.  Contraction of clones, in particular, often occurs once the antigenic stimulus is withdrawn, or if TCR binding is ineffective against the dominant epitopes present in the microenvironment. Identifying expanding and contracting clones has taken different forms in the past, though a popular approach places the problem within a pairwise comparison and multiple testing framework.  One identifies significant changes in clonal proportions between two consecutive time points within a Beta-Binomial testing setting which effectively moderates estimated proportion changes by pulling clones with low read counts toward the null hypothesis of no change \parencite{rytlewski_model_2019}.  The procedure then applies false discovery proportion control to address the severe multiple testing burden involved in the analysis \parencite{benjamini_controlling_1995}.  The approach has been successful at identifying expanding and contracting clones and relating those dynamics to patient intervention and phenotype.  Cancer has been a disease of primary interest with respect to longitudinal TCR clonal movement and their associations with treatment and prognosis, with recent studies demonstrating clonal expansions and contractions significantly associated with radiation therapy directed at metastases \parencite{tang_addition_2023}.  

% The insight motivates hypotheses on the release of neoantigens and influence on 

However, routine implementation of TCR sequencing in clinical practice and management of solid tumors has been limited to date, and particular challenges include heterogeneity and uncertainties in analysis approaches. Comprehensive statistical models for longitudinal T-cell receptor clonal dynamics have so far not been developed in the literature to our knowledge, which might account for variable follow-up by observation, missingness, and error term specification so that what constitutes significant clonal frequency change is explicit.  Several challenges arise when trying to develop a model for these dynamics.  Because of heterogeneity in the total number of template reads within the biological unit being interrogated, one models a fraction, rather than an absolute count, since it is the clonal template read count relative to others that elucidates biological fluctuation and therefore interest.  Second, because of extreme antigenic and therefore receptor diversity, most clones are not ``public'', or shared, across the population being modeled, unlike expression of a certain gene or many gene mutations which though varying across observations have a common interpretation in a sample \parencite{becher_public_2020}.  Therefore one generally cannot relate any one clone and its expansion or contraction with patient phenotype; rather, one must consider dynamics of the repertoire as a whole when modeling patient outcomes.  The challenge of relating clonal dynamics with phenotype is particularly profound for systemic circulating T cells versus T cells that have infiltrated tumor tissue, although the dynamics and functionality of the systemic TCR repertoire are often of considerable clinical interest, and also more easily studied through peripheral blood draws as compared with serial tissue biopsies. Lastly, while it is often feasible to identify change in some proportion from a statistical perspective, identifying biologically relevant changes, which presumably are of greater interest for prognostic models and discovering new biology, is a different and more nuanced question.  Inclusion of an error term that is monotone increasing and approximately scales with the clone's mean parameter may capture biological mechanism best.

% Our approach consists of development of a variance component mixture model, where one component describes the dynamic clones, whereas the static component envisions sampling a clonal mean frequency from some underlying distribution which then parameterizes and constrains the entirety of that clone's frequencies over follow-up. 

We try to address these challenges with an interesting and novel variance component mixture model, which leverages Bayesian conjugacy under two different parameterizations of a variance component model, one component corresponding to dynamic (expanding or contracting) clones, and one component corresponding to static clones \parencite{murphy_machine_2012,searle_variance_2009}.  To give an overview of the model, it posits a clonal mean frequency $\lambda$ (to be indexed according to the dynamic or static parameterization) being sampled hierarchically from a Gamma prior 
% on hyper parameters $\alpha$ and $\beta$, 
which we condition on to subsequently sample from a Poisson distribution on that mean for the read count of the clone, with offset term corresponding to the total number of reads within the biological unit being modeled.  Under the dynamic component, a different clonal mean parameter $\lambda$ is hypothesized for each longitudinal follow-up, giving flexibility to the within-clone frequency variability over time.  In contrast, the same $\lambda$ parameterizes all follow-up times under the static component, constraining the entirety of that clone's frequencies for the times observed.  For both parameterizations, we marginalize out $\lambda$, using conjugacy to arrive at a marginal likelihood that is a product of negative binomial probability mass functions (pmf) for the dynamic component and the negative multinomial pmf for the static component, both of which are modified by the offset.  We fit the $\alpha$ and $\beta$ parameters governing the Gamma prior and $\pi$ specifying the mixing proportion with empirical Bayes, and use the expectation-maximization (EM) algorithm to estimate component membership for each clone, be it dynamic or static, iterating until convergence \parencite{dempster_maximum_1977}.  

In Section \ref{sec:methods}, we derive the model's mixing components, examine scenarios under which each will evaluate better or worse, and develop an EM algorithm for the model's fitting.  In Section \ref{sec:simulation}, we examine sensitivity of model estimation and component membership to different combinations of parameters and follow-up.  In Section \ref{sec:results}, we analyze a cohort of prostate cancer patients randomized to metastasis-directed therapy or not and the influence of that intervention on TCR clonal dynamics.  We conclude in Section \ref{sec:discussion} with consideration of model extensions and improvement.

% we explore several post-hoc measures of clonal dynamics based on model fits with patients outcomes and exposures and identify markers of prognostic value. 

% We perform several posthoc analyses relating measures of dynamics to patients outcomes.  
% peptide chains consisting of     diverse set of receptors expressed on them Many different and coordinating immune cells, are involved The T-cell receptors

\section{Methods}
\label{sec:methods}

% \subsection{Motivation}

Finite mixture models have played a central role in the development of statistical modeling with broad use across data-generating mechanisms where there is belief that subsets of observations arise under a discrete set of different parameters \parencite{fruhwirth-schnatter_finite_2006,celeux_handbook_2018,delmar_mixture_2005}.  Since parameters are generally unknown for each subset, or component, nor is component membership known among observations, these models can be difficult to fit and considerable work has been devoted to understanding how to best do so and behavior of subsequent estimates.  Simulation-based methods like Markov chain Monte Carlo (MCMC) and other optimization routines have proved valuable tools in model fitting
% , and subsequent estimates 
\parencite{nityasuddhi_asymptotic_2003,robert1999monte}.  It may be most common to see mixture models mixing on location and scale parameters in exponential families of distributions for clustering data, though more sophisticated mixtures of regressions or mixtures of experts have also been proposed and applied in different modeling settings especially recently \parencite{fruhwirth-schnatter_model_2019,gormley_mixtures_2018}.  It is less common to mix on the parameterization of a model, which is our proposition here, nor for distinct components to be governed by the same set of hyperparameters, and so the approach distinguishes itself within mixture modeling and presents challenges in turn.  Additionally, while variance components are typically used in mixed models to account for within-observation correlation and augment fixed effects, here we show that they can be useful in partitioning dynamic TCR clones from static ones, that is those that exhibit significant heterogeneity in clonal proportion across time versus those that do not \parencite{searle_variance_2009,yirga_negative_2020,aitkin_general_1999}. The proposition therefore has multiple unique elements in its formulation and has opportunity to provide distinct insight into dynamic and static TCR clone partitioning.

% The two Gamma-Poisson hierarchies upon which our model depends....

% Our mixture consists of 

% The two Gamma-Poisson hierarchies which upon marginalization of the static component yields a negative multinomial distribution governed by two hyperparameters $\alpha$ and $\beta$, whereas marginalization of the dynamic component yields a product of negative binomial distributions governed by the same parameters, 

% where the number of terms is the follow-up times within TCR clonotype.  The arguments for both of these variance components are the number of clonotype TCR reads and the total number of reads across clones per person-time.  

% REMOVE?:

% We fit the model by first randomly assigning a component membership indicator indexed by clone and person $D_{ij}$, optimizing with respect to the $\alpha$ and $\beta$ hyperparameters, calculate component membership expectation $E[D_{ij}]$ according to updated hyperparameters, and iterate until convergence, defined as the mean change in component probabilities being less than some $\epsilon$, here $10^{-8}$.  

% because  The mixing over variance components 

% We arrive at a mixture of variance components, with the dynamic clones evaluates unders a negative multinomial, and a 
% \cite{searle_variance_2009}
% \cite{aitkin_general_1999}
% \cite{yirga_negative_2020}

\subsection{Model framework}

Consider the $i^{th}$ clone template read count measured with immunosequencing for person $j$ at time $k$, $C_{ijk}$, and the total number of template counts for person $j$ at time $k$, $O_{jk} = \sum_{i=1}^{U_{jk}} C_{ijk}$, where $U_{jk}$ is the number of unique clones for person $j$ at time $k$.  
% 
% In so doing, one can relate their dynamics to patient exposures and outcomes.  
% 
% REMOVE: with $j\in \{1,\dots , T_i\}$, where some clones are not fully observed and observed at 2 of three points in the index set. 
% 
% \vspace{0.3cm}
% \noindent
We seek to distinguish the dynamic clones (denoted $D_{ij}=1$ for clone $i$ person $j$), who exhibit significant contraction or expansion over longitudinal follow-up, from the static ones ($D_{ij}=0$). We assume $D_{ij}\sim \mbox{Bernoulli}(\pi)$ for all $i$ and $j$. Since there is variability in the total number of template reads and it is of primary biological interest how proportions of clones change over time, variation in the proportion $C_{ijk}/O_{jk}$ will play a central role in a clone's evaluation into the dynamic or static component of the mixture model.  
% and relate the numbers and absolute movement of such clones to patient intervention, disease state, prognosis, and biomarkers at baseline and follow-up. 
Specifically, the hierarchical formulation envisions that among the static component, clonal mean parameter $\lambda_{ij} \sim \mbox{Gamma}(\alpha,\beta)$  is invariant over time $k$ and sampled from a Gamma distribution on hyperparameters $\alpha$ and $\beta$.  Template count reads are assumed Poisson-distributed $C_{ijk} \sim \mbox{Pois}(\lambda_{ij} O_{jk})$, where the mean parameter $\lambda_{ij}$ is scaled by the total number of reads $O_{jk}$ for person $j$ at time $k$, which yields an interpretation of that mean parameter on the proportion scale.
% implicitly modeling clonal proportion dynamics.
% , effectively treating its logged value as an offset.  
In contrast, under the dynamic component model, separate $\lambda_{ijk}$ are assumed sampled for each time point, and in which case $C_{ijk}$ is subsequently sampled according to $C_{ijk} \sim \mbox{Pois}(\lambda_{ijk} O_{jk})$.  Integration out of $\lambda_{ij}$ and $\lambda_{ijk}$ in these two Gamma-Poisson hierarchies yields respective posterior predictive distributions, which due to conjugacy are the negative multinomial and product of negative binomial distributions, where the number of terms for the latter is the follow-up times within TCR clonotype.  The arguments for evaluating these distributions are the number of clonotype TCR reads and the total number of reads across clones per person-time.  In contrast to modeling the clonal template counts with only a Poisson distribution, the extra degree of freedom accorded by the Gamma prior more flexibly accounts for the right-skewness in the TCR template read counts.  

% because of the right-skewed and significant dynamic ranges of this outcome. 

% accordingly, each a function of the same hyperparameters $\alpha$ and $\beta$.  
% a product of negative binomial distributions governed by the same parameters, 

% These distributions are appropriate for modeling clonal template counts because of the right-skewed and significant dynamic ranges of this outcome.  
% These distributions are modified according to the $O_{jk}$ via their use as an offset, implicitly modeling clonal proportion dynamics.  

% the variable number of total person-time template reads

% In Section \ref{} we develop the model, in section \ref{} demonstrate it in simulation and characterize its behavior different dynamics and read characteristics.  We conclude with an application to on longitudinal TCR study using radiation and progression free survival.  We give thoughts on extensions in section \ref{discussion}.  

% , which normalize for variable total number of reads and for which is a common means of modelling (there is precedent in the literature).  

% frequencies are distributed negative binomial, whose counts in this case we envision as the number of (0.001) units of the frequency.  
% % , that is scale the frequency by some number.  
% Because nearly all counts have well below 1\% frequency, with the maximum frequency only 3-5\% in a small subset of cases, we do not saturate against the theoretical ceiling of 1
% % since these are frequencies 
% and so can conceptually think of the quantity as unbounded from above, consistent with the negative binomial parameterization. 

\subsection{Static model}

The derivation of the static component clonal model then can be written as 

$$C_{ijk} \,|\, (\lambda_{ij}, O_{jk}, D_{ij}=0) \sim \mbox{Pois}(\lambda_{ij} O_{jk} )$$

with 

$$\lambda_{ij} | (D_{ij}=0) \sim \mbox{Gamma}(\alpha, \beta)$$

The likelihood for $C_{ij \cdot} =(C_{ij1} \dots C_{ijT_j})$ given $\lambda_{ij}$ and $O_{j \cdot} =(O_{j1} \dots O_{jT_j})$, $l(C_{ij\cdot} | \lambda_{ij}, O_{j\cdot })=\prod_k l(C_{ijk} | \lambda_{ij}, O_{jk })$, with $T_j$ the number of time points observed for clone $i$ on person $j$ due to an unbalanced design (ie, missingness of follow-up for a subset of patients), is

$$l(C_{ij\cdot} | \lambda_{ij} ,  O_{j\cdot }) = \prod_{k=1}^{T_j} \frac{\exp(-\lambda_{ij}O_{jk})\, (\lambda_{ij} O_{jk})^{C_{ijk}}}{C_{ijk}!} = \frac{\exp(- \lambda_{ij} \sum_k O_{jk}) \lambda_{ij}^{\sum_k C_{ijk}} \prod_{k=1}^{T_j} O_{jk}^{C_{ijk}}}{\prod_{k=1}^{T_j} C_{ijk}!}$$

Since our prior on $\lambda_{ij}$, $p(\lambda_{ij} | D_{ij}=0, \alpha, \beta)$, is Gamma$(\alpha,\beta)$, the conditional joint distribution of $(C_{ij\cdot} , \lambda_{ij})\, |\, O_{j\cdot}, D_{ij}=0$ is 

\begin{align}
p(C_{ij\cdot} , \lambda_{ij}& \,|\, O_{j\cdot}, D_{ij} =0, \alpha, \beta) = l(C_{ij\cdot} | \lambda_{ij}, \, O_{j\cdot})p(\lambda_{ij} | D_{ij} = 0, \alpha, \beta) \nonumber \\
=&\frac{\exp(- \lambda_{ij} \sum_k O_{jk}) \lambda_{ij}^{\sum_{k} C_{ijk}} \prod_{k=1}^{T_j} O_{jk}^{C_{ijk}} }{\prod_{k=1}^{T_j} C_{ijk}!} \cdot \frac{\beta^{\alpha}}{\Gamma(\alpha)} \lambda_{ij}^{\alpha-1} \exp{(-\beta \lambda_{ij})}
\label{eqn:first}
\end{align}
Collecting terms of Equation (\ref{eqn:first}) and leveraging conjugacy, we recognize the kernel of the Gamma on updated parameters $\alpha'= \sum_k C_{ijk} + \alpha $ and $\beta'= \sum_k O_{jk} + \beta$.  Then integrating out $\lambda_{ij}$, we obtain the inverse normalizing constant of the Gamma($\alpha',\beta'$)

\begin{align}
p(C_{ij\cdot } |O_{j\cdot}, D_{ij}=0, \alpha, \beta)=&\int p(C_{ij\cdot },\lambda_{ij} | O_{j\cdot }, D_{ij}=0, \alpha, \beta) \, d\lambda_{ij} \nonumber \\
=& \frac{\beta^{\alpha} \,  \prod_{k=1}^{T_j} O_{jk}^{C_{ijk}}}{ \Gamma(\alpha) \, \prod_{k=1}^{T_j} C_{ijk}! }  \int_0^\infty  \lambda_{ij}^{\sum_k C_{ijk} + \alpha-1} \exp{(-\lambda_{ij}(\beta+ \sum_k O_{jk}))} d\lambda_{ij} \nonumber \\ 
 =& \frac{\beta^{\alpha} \,  \prod_{k=1}^{T_j} O_{jk}^{C_{ijk}}}{ \Gamma(\alpha) \, \prod_{k=1}^{T_j} C_{ijk}!} \cdot \frac{ \Gamma(\sum_k C_{ijk} + \alpha)}{(\beta + \sum_k O_{jk})^{\sum_k C_{ijk} + \alpha}}  \nonumber \\ % \cdot  \int_0^\infty  \lambda_i^{\sum_j C_{ij} + \alpha-1} \exp^{-\lambda_i(\beta_S  + T_i)} d\lambda_i$$ 
=&\frac{ \Gamma(\sum_k C_{ijk} + \alpha) }{ \Gamma(\alpha) \, \prod_{k=1}^{T_j} C_{ijk}!}  \biggl(\frac{ \beta}{\beta + \sum_k O_{jk}}\biggr)^{\alpha}  \prod_{k=1}^{T_j} \biggl( \frac{O_{jk}}{\beta + \sum_k O_{jk}}\biggr)^{C_{ijk} }
\label{eqn:second}
\end{align}
\noindent which for discrete $\alpha$ we recognize Equation (\ref{eqn:second}) as a negative multinomial probability mass function on size parameter $\alpha$, success probability $\beta/(\sum_k O_{jk} +\beta)$, and ``offset-normalized''  failure probabilities $O_{jk}/(\beta + \sum_k O_{jk})$, $k=1, \dots , T_j$, 
% $T_j$ the number of observed follow-ups for observation $j$, 
 and evaluated at vector $(C_{ij1},\dots ,C_{ijT_j})$,
 for clone read counts $C_{ij\cdot}$, noting $\Gamma(n) = (n-1)!$ for integers $n$.  When the $O_{jk}$ are invariant to $k$, the failure probabilities will be equal-valued.  Regardless of dependence on $k$, the form of the expression suggests that it will tend to evaluate best when vector $(C_{ij1},\dots ,C_{ijT_j})$ is approximately proportional to  $(O_{j1},\dots ,O_{jT_j})$, the interpretation of which is that $C_{ijk}/O_{jk}$ is approximately constant and therefore expected for this static component model.  

 % $k=\alpha$, success probability $\beta/(\beta + \sum_k O_{jk})$, and ``offset-normalized'' equal valued failure probabilities $O_{jk}/(\beta + \sum_k O_{jk})$,..... the number of observed follow-ups of person j for all their clones $C_{ij\cdot}$, and evaluated at vector $(C_{ij1},\dots ,C_{ijT_j}, \alpha_S-1)$, noting $\Gamma(n) = (n-1)!$ for integers $n$.

\subsection{Dynamic Model}

The dynamic component clonal model can be derived with

$$C_{ijk} \,|\, (\lambda_{ijk}, O_{jk}, D_{ij}=1) \sim \mbox{Pois}(\lambda_{ijk} O_{jk} )$$

where

$$\lambda_{ijk} | (D_{ij}=1) \sim \mbox{Gamma}(\alpha, \beta)$$

% ``pi AGAIN DOESN'T SHOW UP HERE BC WE CONDITION ON Dij''

We remark that while $\lambda_{ij}$ in the static model can be considered a random effect and has an expectation which is identifiable, here $\lambda_{ijk}$ varies with time $k$ and so is non-identifiable and comparison of the dynamic and static components is only possible under marginalization and the induced posterior predictive distributions.  In this case, integration out of $\lambda_{ijk}$ yields:
% a product of negative binomial distributions:

\begin{align}
p(C_{ij\cdot}\,  |\, O_{j\cdot} ,D_{ij}=1, & \alpha, \beta) =  \prod_{k=1}^{T_j} \int p(C_{ijk },\lambda_{ijk} | O_{j \cdot}, D_{ij}=1, \alpha, \beta) \, \,  d\lambda_{ijk}   \nonumber \\ 
= & \prod_{k=1}^{T_j} \int l(C_{ijk} | \lambda_{ijk}, \, O_{j\cdot})p(\lambda_{ijk}  | D_{ij} = 1, \alpha, \beta) \, d\lambda_{ijk} \nonumber \\ 
 % p(C_{ij\cdot} , \lambda_{ij}& \,|\, O_{j\cdot}, D_{ij} =0, \alpha, \beta)  \nonumber \\
= &  \prod_{k=1}^{T_j} \int \frac{\exp(- \lambda_{ijk} \, O_{jk}) \lambda_{ijk}^{ C_{ijk}}  O_{jk}^{C_{ijk}} }{ C_{ijk}!} \cdot \frac{\beta^{\alpha}}{\Gamma(\alpha)} \lambda_{ijk}^{\alpha-1} \exp{(-\beta \lambda_{ijk})} 
\,  d\lambda_{ijk}  \nonumber \\
% =  & \int \frac{\exp(- \sum_k \lambda_{ijk} ( O_{jk} + \beta) ) \prod_{k=1}^{T_j} O_{jk}^{C_{ijk}} \, \lambda_{ijk}^{ C_{ijk}+\alpha-1}  }{\prod_{k=1}^{T_j} C_{ijk}!} \cdot \frac{\beta^{\alpha}}{\Gamma(\alpha)}  \,  d\lambda_{ijk}   \\
% 
% & = \prod_{k=1}^{T_j} p(C_{ijk}| O_{j\cdot} , D_{ij}=1, \alpha, \beta) \\ 
= & \prod_{k=1}^{T_j} \frac{ \Gamma( C_{ijk} + \alpha) }{ \Gamma(\alpha)  C_{ijk}!}  \biggl(\frac{ \beta}{\beta + O_{jk}}\biggr)^{\alpha}   \biggl( \frac{O_{jk}}{\beta + O_{jk}}\biggr)^{ C_{ijk}} \label{eqn:dyn}
\end{align}
where we write the distribution in terms of the $\Gamma(\cdot)$ function, and which in $C_{ijk}$ we recognize as the product of $T_j$ negative binomial mass functions on size parameter $\alpha$ and success probabilities $\beta/(\beta + O_{jk})$ for $k=1,\dots , T_j$.  We develop our mixture model using these mass functions in Equations (\ref{eqn:second}) and (\ref{eqn:dyn}) for the static and dynamic components, respectively.
% Note that while random effect $\lambda_{ij}$ in the static model is identifiable and whose expectation could be calculated, $\lambda_{ijk}$ is not, and comparison of the two components is only possible under marginalization and the induced posterior predictive distributions.  

% So conditional on $\alpha$ and $\beta$ and using Equations (\ref{eqn:first}) and (\ref{eqn:second}), we can calculate 

% and note $\Gamma(n) = (n-1)!$
%}{(\beta_S + T_i)^{\sum_j C_{ij} + \alpha_S}} $$ 

% evaluation of the two models identifiable in the model and only comparable with the static component model via marginalization over $\lambda_{ijk}$ and evaluation under the induced posterior predictive negative binomial distribution.

% we recognize as the product of modified negative binomials.

\subsection{Comparing Components}

To consider the circumstances under which a $C_{ij\cdot}$ vector will evaluate more favorably under the static or dynamic model, one can simplify their quotient and examine terms.  Taking the static over the dynamic model, that is Equation (\ref{eqn:second}) over (\ref{eqn:dyn}), yields a reduced expression for which static behavior would be associated with higher values:

% static:
% $$\frac{ \Gamma(\sum_k C_{ijk} + \alpha) }{ \Gamma(\alpha) \prod_{k=1}^{T_j} C_{ijk}!}  \biggl(\frac{ \beta}{\beta + \sum_k O_{jk}}\biggr)^{\alpha}  \prod_{k=1}^{T_j} \biggl( \frac{O_{jk}}{\beta + \sum_k O_{jk}}\biggr)^{C_{ijk}}$$
%  % \prod_{k=1}^{T_j} 
% dynamic
% $$\prod_{k=1}^{T_j} \frac{ \Gamma( C_{ijk} + \alpha) }{ \Gamma(\alpha)\, C_{ijk}!}  \biggl(\frac{ \beta}{\beta + O_{jk}}\biggr)^{\alpha}   \biggl( \frac{O_{jk}}{\beta + O_{jk}}\biggr)^{ C_{ijk}}$$
\begin{align}
\frac{p(C_{ij\cdot}\,  |\, O_{j\cdot} ,D_{ij}=0, \alpha, \beta)}{p(C_{ij\cdot}\,  |\, O_{j\cdot} ,D_{ij}=1, \alpha, \beta)} = \kappa_1 \cdot \frac{\Gamma(\sum_{k=1}^{T_j} C_{ijk} + \alpha)}{ \prod_{k=1}^{T_j} \Gamma(C_{ijk} + \alpha)}\cdot   \prod_{k=1}^{T_j} \biggl( \frac{\beta +  O_{jk}}{\beta + \sum_k O_{jk}}\biggr)^{C_{ijk} }
\label{eqn:quotient}
\end{align}
where the proportionality constant $\kappa_1$ is a function of $\alpha$, $\beta$, and $O_{j\cdot}$, that is
$$\kappa_1 = \Gamma(\alpha)^{T_j-1} \bigg(\frac{\beta^{(1-T_j)} \prod_{k=1}^{T_j} (\beta + O_{jk} ) }{\beta + \sum_{k=1}^{T_j} O_{jk} }\bigg)^\alpha $$
We can multiply and divide by $\rho^{\sum_k C_{ijk}}$, and add and subtract $\alpha-1$ in the exponent of each of the $T_j$ terms and re-write the expression (\ref{eqn:quotient}) as 
\begin{align}
 = \kappa_1 \kappa_2 \cdot \frac{\Gamma(\sum_{k=1}^{T_j} C_{ijk} + \alpha)}{ \prod_{k=1}^{T_j} \Gamma(C_{ijk} + \alpha)}\cdot \prod_{k=1}^{T_j} \biggl(\frac{\rho (\beta + O_{jk})}{\beta + \sum_k O_{jk}}\biggr)^{C_{ijk} +\alpha-1}
\label{eqn:multinom}
 \end{align}
where $\rho$ defined as follows allows the product terms to be interpreted as multinomial probabilities and $\kappa_2$ is a normalizing constant, with $$\rho= \bigg[ \sum_{k=1}^{T_j} \biggl( \frac{\beta +  O_{jk}}{\beta + \sum_k O_{jk}}\biggr) \bigg]^{-1} \;\;\;\;\;\; \mbox{and}\;\;\;\;\;\;\; \kappa_2=  \rho^{-\sum_k (C_{ijk}+\alpha-1)} \prod_{k=1}^{T_j} \biggl( \frac{\beta +  O_{jk}}{\beta + \sum_k O_{jk}}\biggr)^{-\alpha +1}$$
One observes that $\rho$ is a function of $\beta$ and $O_{j\cdot}$, and $\kappa_1 \kappa_2$ is a function of $O_{j\cdot}, \sum_k C_{ijk}, \alpha,\beta$, and therefore constant for changes in $C_{ij\cdot}$ holding $\sum_k C_{ijk}$ constant.

If we conceptualize expression (\ref{eqn:multinom}) as conditioning on the $O_{j\cdot}$  and $ \sum_k C_{ijk} $, this is multinomial and will evaluate highest with $C_{ijk}$ as roughly proportional to $\beta + O_{jk}$ then offset by $\alpha$.  Assuming $O_{jk}$ dominates $\beta$ and $C_{ijk}$ dominates $\alpha$ as is roughly the case in practice with low expected frequencies relative to high variation, the mode will be achieved when $C_{ijk}$ is approximately proportional to $O_{jk}$.  This implies that $C_{ijk}/O_{jk}$ is approximately constant and hence the quotient evaluates to a relatively high value up to the normalizing constant, meaning the static component is favorable as compared to the dynamic component.  In contrast, the expression will be minimized when in the tails of the multinomial, that is, when vector $C_{ij\cdot}$ is far from proportionality with $O_{jk}$.  But holding $\sum_k C_{ijk}$ constant, this entails some $C_{ijk}$'s to be small and some large, which is to say the $C_{ijk}$ exhibit dynamic and highly variable behavior.

\subsection{Mixing and Model fitting}
% ``BY CONDITION INDEPENDENT OF Dij and and alpha beta given Cij (consider `plate' notation of wikipedia mixture model page) '' 
Having derived component expressions, we can write the likelihood in terms of them and the component indicator Bernoulli $D_{ij}$ where its distribution is described with
$p(D_{ij}=1 | \pi) = \pi$.  We write 
\begin{align*}
p(C_{ij\cdot}\,  |\, O_{j\cdot}  , \alpha, \beta, \pi)  &=p(C_{ij\cdot}\, ,   D_{ij}=1 |  \, O_{j\cdot} , \alpha, \beta, \pi) + p(C_{ij\cdot}\, , D_{ij}=0 |\, O_{j\cdot} , \alpha, \beta, \pi) \\
=p(C_{ij\cdot}\, & |\, O_{j\cdot} ,D_{ij}=1,  \alpha, \beta) p(D_{ij}=1 | \pi) + p(C_{ij\cdot}\,  |\, O_{j\cdot} ,D_{ij}=0, \alpha, \beta) p(D_{ij}=0 | \pi) \\ 
= \pi \cdot p(& C_{ij\cdot}  \,  |\, O_{j\cdot} ,D_{ij}=1, \alpha, \beta) + (1-\pi) \cdot p(C_{ij\cdot}\,  |\, O_{j\cdot} ,D_{ij}=0, \alpha, \beta) 
\end{align*}
since $p(D_{ij}\,|\, \pi)$ is not a function of $\alpha$ nor $\beta$.
% \textcite{dempster_maximum_1977}
% (NEED TO MOVE THIS TO A MORE GENERAL PART BC NOT SPECIFIC TO STATIC)
% ``SINCE WE'RE CONDITIONONG ON Dij in what FOLLOWS, PI WON'T SHOW UP FOR A FEW PARAGRAPHS''
% 
% After an initial random cluster membership assignment of clones, we iterate between ma
% 
Hyperparameters $\alpha$ and $\beta$ are estimated using empirical Bayes and the model is fit using the EM algorithm \parencite{dempster_maximum_1977}.  We calculate the expectation of the complete data log-likelihood
\begin{align*}
\sum_{ijk} \log p(C_{ijk} | O_{jk} , \alpha, \beta , \pi) = \sum_{ijk} \log\big( p( & C_{ijk} ,D_{ij}=1  |   O_{jk} , \alpha, \beta , \pi )  \\
& \hspace{0.1cm}+  p(C_{ijk},D_{ij}=0 | O_{jk} , \alpha, \beta , \pi ) \, \big) 
%  = \sum_{ijk} \log\big( p(C_{ijk}  | &  O_{jk} ,D_{ij}=1, \alpha, \beta )p(D_{ij}=1 | \pi)  \\
% & \hspace{0.1cm}+  p(C_{ijk} | O_{jk} ,D_{ij}=0, \alpha, \beta )p(D_{ij}=0 | \pi) \, \big) \\
 % = \sum_{ijk} \log\big( p(C_{ijk}  | &  O_{jk} ,D_{ij}=1, \alpha, \beta )p(D_{ij}=1 | \alpha, \beta)  \\
% & \hspace{1cm}+  p(C_{ijk} | O_{jk} ,D_{ij}=0, \alpha, \beta )p(D_{ij}=0 | \alpha, \beta) \big) \\ 
%  = \sum_{ijk} \log\big(  \pi \cdot p(&   C_{ijk}  |  O_{jk} ,D_{ij}=1, \alpha, \beta )  \\
% & \hspace{0.1cm}+ (1-\pi)\cdot  \; p(C_{ijk} | O_{jk} ,D_{ij}=0, \alpha, \beta ) \, \big)
\end{align*}
for the $(m+1)^{st}$ iteration under $p(D_{ij} |C_{i j \cdot}  ,O_{ j \cdot}, \alpha^{(m)},  \beta^{(m)},\pi^{(m)})$, the distribution for latent variables $D_{ij}$ given current estimates $(\alpha^{(m)}, \beta^{(m)},\pi^{(m)})$.  This distribution for the latent variables given the data is 
% where an expression for this distribution is given with 

% E[D_{ij} | C_{i j \cdot},O_{ j \cdot},&  \alpha^{(m)},\beta^{(m)},\pi^{(m)}] = 

\begin{align}
p(D_{ij} |C_{i j \cdot} & ,O_{ j \cdot}, \alpha^{(m)},  \beta^{(m)},\pi^{(m)}) \nonumber \\
= &\frac{p(D_{ij} | C_{i j \cdot},O_{ j \cdot}, \alpha^{(m)}, \beta^{(m)}, \pi^{(m)}) }{p(D_{ij}=1 | C_{i j \cdot},O_{ j \cdot}, \alpha^{(m)}, \beta^{(m)}, \pi^{(m)})  + p(D_{ij}=0 | C_{i j \cdot},O_{ j \cdot}, \alpha^{(m)}, \beta^{(m)},\pi^{(m)}) } \nonumber  \\
= &\frac{p(D_{ij}, C_{i j \cdot} |O_{ j \cdot}, \alpha^{(m)}, \beta^{(m)}, \pi^{(m)}) }{p(D_{ij}=1 , C_{i j \cdot} | O_{ j \cdot}, \alpha^{(m)}, \beta^{(m)}, \pi^{(m)})  + p(D_{ij}=0 , C_{i j \cdot} | O_{ j \cdot}, \alpha^{(m)}, \beta^{(m)},\pi^{(m)}) }  \label{eqn:prop} \\
= &\frac{\pi^{(m)}  \; p(C_{ij \cdot} | O_{j\cdot } ,D_{ij}, \alpha^{(m)}, \beta^{(m)} )}{
\pi^{(m)} \; p(C_{ij \cdot } | O_{j \cdot } ,D_{ij}=1, \alpha^{(m)}, \beta^{(m)} )
+ (1-\pi^{(m)})  \; p(C_{ij \cdot } | O_{j \cdot } ,D_{ij}=0, \alpha^{(m)}, \beta^{(m)} )} \label{eqn:em_other}  
% = &  \, p(D_{ij}=1 |   C_{ij \cdot} , O_{j\cdot } , \alpha^{(m)}, \beta^{(m)} )
\end{align}
where Equation (\ref{eqn:prop}) uses $p(D_{ij} | C_{i j \cdot},O_{ j \cdot}, \alpha^{(m)}, \beta^{(m)}, \pi^{(m)}) \propto p(D_{ij} , C_{i j \cdot} | O_{ j \cdot}, \alpha^{(m)}, \beta^{(m)}, \pi^{(m)})$ and (\ref{eqn:em_other}) uses $p(D_{ij} , C_{i j \cdot} | O_{ j \cdot}, \alpha^{(m)}, \beta^{(m)}, \pi^{(m)}) = \pi^{(m)D_{ij}}(1 - \pi^{(m)})^{1-D_{ij}}  \; p(C_{i j \cdot} | O_{j \cdot } ,D_{ij}, \alpha^{(m)}, \beta^{(m)} )$ and is calculated with component expressions (\ref{eqn:second}) and (\ref{eqn:dyn}).

% \begin{align*}
% p(D_{ij} | C_{i j \cdot}, & O_{ j \cdot} , \alpha^{(m)}, \beta^{(m)},\pi^{(m)}) \\ 
% =&\frac{\pi^{(m)} \, p(C_{ij\cdot}\,  |\, O_{j\cdot} ,D_{ij},  \alpha^{(m)}, \beta^{(m)}) }{\pi^{(m)} \, p(C_{ij\cdot}\,  |\, O_{j\cdot} ,D_{ij}=1,  \alpha^{(m)}, \beta^{(m)})  + (1-\pi^{(m)}) \, p(C_{ij\cdot}\,  |\, O_{j\cdot} ,D_{ij}=0,  \alpha^{(m)}, \beta^{(m)})}
% \end{align*}

We then maximize the expectation of the complete data log-likelihood under the latent variable distribution
% $p(D_{ij} |C_{i j \cdot}  ,O_{ j \cdot}, \alpha^{(m)},  \beta^{(m)},\pi^{(m)})$ 
and with respect to $\alpha, \beta, \pi$ to find $\alpha^{(m+1)}, \beta^{(m+1)}$, and $\pi^{(m+1)}$.  That is, we find 
$$Q(\alpha,\beta,\pi | \alpha^{(m)}, \beta^{(m)},\pi^{(m)}) = E_{p_D}[\log\, p(C_{ijk},D_{ij} | O_{jk}, \alpha,\beta,\pi)]$$
where for brevity $p_{D} = p(D_{ij} | C_{ijk},O_{jk}, \alpha^{(m)}, \beta^{(m)}, \pi^{(m)} ) $, and subsequently calculate 
$$({\alpha}^{(m+1)},\; {\beta}^{(m+1)},\;{\pi}^{(m+1)} ) = \mbox{arg\,max}_{\alpha,\beta,\pi} Q(\alpha,\beta,\pi | \alpha^{(m)}, \beta^{(m)}, \pi^{(m)})$$
\noindent
Since $\alpha$ and $\beta$ are present in both components, maximization is difficult.  While an additional, nested EM is possible to estimate  $(\alpha^{(m+1)}, \beta^{(m+1)}, \pi^{(m+1)})$, here we use BFGS for the maximization step \parencite{broyden_convergence_1970,fletcher_practical_2000}.

We fit the model by first randomly assigning a component membership indicator indexed by clone and person $D_{ij}$, optimize with respect to $\alpha$, $\beta$, and $\pi$, and calculate the latent variable distribution given the data and current update of the parameters.  
% component membership expectation $E[D_{ij}]$ according to updated hyperparameters, 
We iterate EM steps until convergence, defined as the mean squared change in component probabilities 
$$\sum_{ij} \big(E[D_{ij} | C_{i j \cdot},O_{ j \cdot}, \alpha^{(m)}, \beta^{(m)}, \pi^{(m)}] - E[D_{ij} | C_{i j \cdot},O_{ j \cdot}, \alpha^{(m+1)}, \beta^{(m+1)}, \pi^{(m+1)}]\big)^2/n $$
% (D_{ij}^{(m)} - D_{ij}^{(m+1)})/n$ 
for iteration $m$ to $m+1$ being less than some $\epsilon$, chosen here as $10^{-8}$.

% mean change in component probabilities being less than some $\epsilon$, here $10^{-8}$.  

% We iterate EM steps until the convergence criterion is reached, the mean of changes $\sum_{ij} (D_{ij}^{(m)} - D_{ij}^{(m+1)})/n$ for iteration $m$ to $m+1$ less than some $\epsilon$, chosen here $10^{-8}$.

\section{Simulation}
\label{sec:simulation}

\subsection{Generating process}

We generated 60,000 clonotypes under a combination of $\alpha$, $\beta$, and  $\pi$ values as shown in Table \ref{tab:est} and with offsets $O_{jk}$ sampled from an exponential distribution that approximated the empirically observed total template read counts for person-times.  Clonotype mean proportions were sampled from a $\mbox{Gamma}(\alpha,\beta)$ for the different parameter combinations, and subsequently multiplied by the $O_{jk}$ to be used as the mean parameter in sampling from a Poisson distribution.  This mean parameter was either varying or constant across the variable number of follow-up times as a function of whether the clone was considered dynamic or static, respectively.  Depending on the simulation scenario, we fit the variance component mixture model to two, three, or four of the generated time points with and without missingness to examine the influence of follow-up on parameter estimation and clonotype classification sensitivity and specificity, with 20\% of clones generated as dynamic and 80\% as static to resemble what was observed in practice.  We iterated 100 times for each parameter combination and number of follow-ups to assess variability in estimation to these dimensions.  

% Parameter estimates for $\hat{\alpha}$, $\hat{\beta}$, and $\hat{\pi}$ were calculated with two, three, and four follow-up periods depending on the scenario.  

% We fit the variance component mixture model to these data and calculated  assuming four followup times, then sensitivity and specificity of the dynamic clones varying the number of followup times.  

\subsection{Simulation Results}

Simulations indicate model fits reflect the parameters under which data are generated.  Parameter estimates demonstrate a relative lack of bias and low variance as shown in Table \ref{tab:est} under variable parameter combinations. Table \ref{tab:estVari} likewise shows relatively little impact of missingness on estimation, and only modest influence of variable follow-up on it.  One observes a slight elevation in bias for both $\hat{\alpha}$ and $\hat{\beta}$ when estimating the parameters based on fewer follow-up periods, though in all cases it is $< 3\%$.  Because both are elevated in the same degree, it has negligible influence on the expected frequency implied by the parameter quotient given the Gamma prior.  Convergence criteria for the different parameter combinations were achieved in fewer than 25 EM iterations.  

\begin{figure}[H]
\begin{subfigure}[h]{0.35\linewidth}
\includegraphics[width=1.2\textwidth]{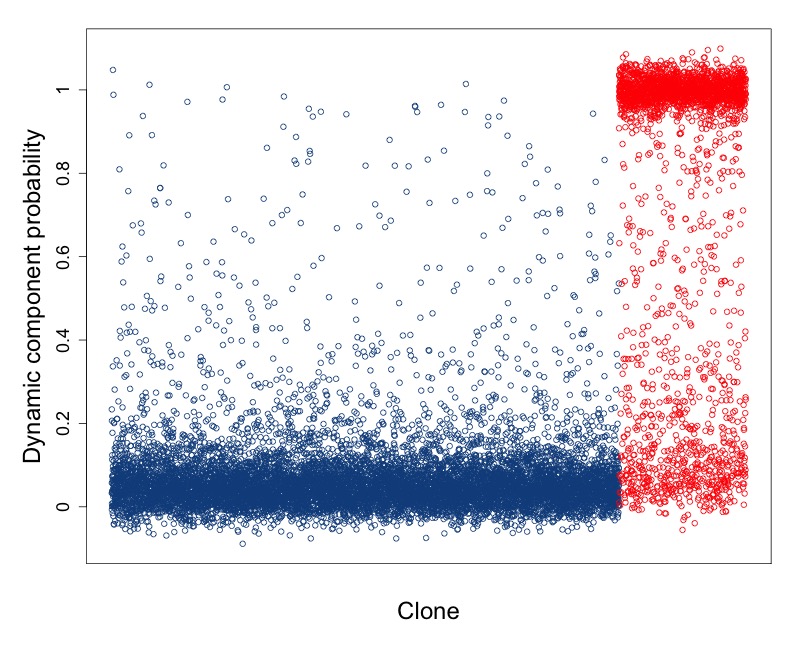} % viz_eff0_7.pdf}
\caption{Clonotype membership probability with two observed follow-ups.}
% \label{fig:sim1}  %70}
\end{subfigure}
% \hfill
\hspace{1cm}
\begin{subfigure}[h]{0.35\linewidth}
\includegraphics[width=1.2\textwidth]{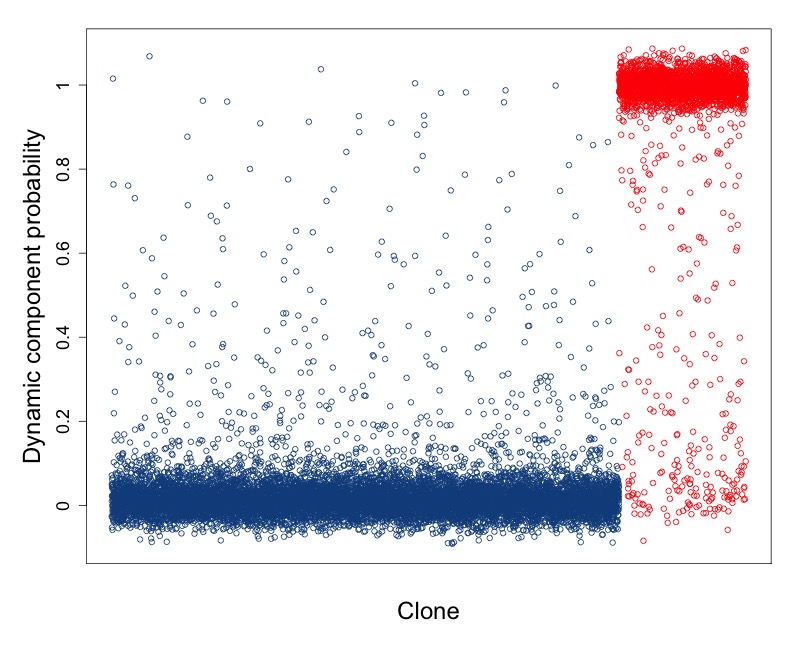} % viz_eff0_7.pdf}
\caption{Clonotype membership probability with three observed follow-ups.}
% \label{fig:sim1}  %70}
\end{subfigure}
\hfill
\newline
% \hspace{3cm}
\centering
\begin{subfigure}[h]{0.35\linewidth}
% \begin{center}
% \centering
\includegraphics[width=1.2\textwidth]{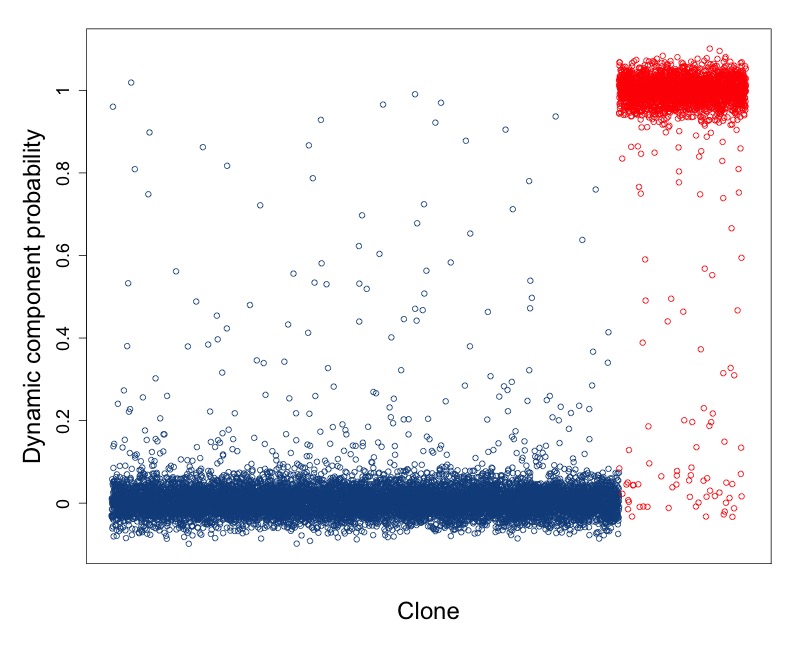} % viz_eff0_7.pdf}
\caption{Clonotype membership probability with four observed follow-ups.}
% \label{fig:sim1}  %70}
% \begin{center}
\end{subfigure}
\caption{Model estimated clonotype membership probability for those generated under the static (blue points) versus dynamic models (red points) with two, three, and four within-clone observed follow-ups and $\alpha=1, \, \beta=200$. Gaussian jitter is added with respect to the y-axis to decrease overplotting.}
\label{fig:dot}
\end{figure}

Table \ref{tab:sens} shows that under the parameters considered, specificity was nearly perfect and relatively invariant to the range of component membership thresholds used of 0.75 to 0.95.  Sensitivity to identifying dynamic clones decreased as a function of the threshold used, whereas it increased significantly in the number of follow-ups, each holding the other variable constant and as expected.  The results indicate that the more information one has within clone longitudinally, the greater the ability to distinguish between dynamic and static behavior. This behavior is demonstrated visually in Figure \ref{fig:dot}, as one can see more dense dynamic and static estimated clone probabilities near the top and bottom, respectively, of the figures as the number of modeled follow-ups increases.  The pattern suggests less misclassification under any threshold as longitudinal information accumulates.

% Table \ref{tab:miss} shows missingness.....

 % latex table generated in R 4.2.1 by xtable 1.8-4 package
 % Sat Aug 10 23:17:02 2024
 \begin{table}[ht]
 \centering
 \begin{tabular}{ccclll}
   \hline
   \multicolumn{3}{c}{Parameters:}  & & & \\
  %  \hline
  % $\alpha$ & $\beta$ & $\pi$& $\hat{\alpha}\mbox{ with }95\%\, \mbox{CI}\,\, (\alpha=1\mbox{ or }2)$ & $\hat{\beta} \mbox{ with }95\%\, \mbox{CI}\,\,\, (\beta=100\mbox{ or }200)$ & $\hat{\pi} \mbox{ with }95\%\, \mbox{CI}\,\,\, (\pi=0.2)$ \\
% \hline
  $\alpha$ & $\beta$ & $\pi$& $\hat{\alpha}\mbox{   (}95\%\, \mbox{CI)}$ & $\hat{\beta} \mbox{   (}95\%\, \mbox{CI)}$ & $\hat{\pi} \mbox{   (}95\%\, \mbox{CI)}$ \\
   \hline
 1 &100 &0.2 & 1.01 (1.01,1.02) & 101.25 (100.32,102.41) & 0.2 (0.2,0.2) \\
  1  &200 &0.2 & 1.02 (1.02,1.03) & 204.04 (202.11,206.48) & 0.21 (0.2,0.21) \\
  2 &100 & 0.2 & 2.01 (1.99,2.02) & 100.34 (99.28,101.33) & 0.2 (0.2,0.2) \\
  2 & 200 & 0.2  & 2.02 (2.00,2.03) & 200.84 (199.19,203.23) & 0.2 (0.2,0.2) \\
    \hline
 \end{tabular}
 \caption{Component and mixing parameter estimates and confidence intervals under four combinations of $\alpha$ and $\beta$ based on 100 iterations of each scenario.}
\label{tab:est}
\end{table}

%  % latex table generated in R 4.2.1 by xtable 1.8-4 package
%  % Sat Aug 10 23:17:02 2024
%  \begin{table}[ht]
%  \centering
%  \begin{tabular}{clll}
%    \hline
%    % \multicolumn{1}{c}{Scenario:}  & & & \\
%   %  \hline
%   % $\alpha$ & $\beta$ & $\pi$& $\hat{\alpha}\mbox{ with }95\%\, \mbox{CI}\,\, (\alpha=1\mbox{ or }2)$ & $\hat{\beta} \mbox{ with }95\%\, \mbox{CI}\,\,\, (\beta=100\mbox{ or }200)$ & $\hat{\pi} \mbox{ with }95\%\, \mbox{CI}\,\,\, (\pi=0.2)$ \\
% % \hline
%    Follow-up scenario & $\hat{\alpha}\mbox{   (}95\%\, \mbox{CI)}$ & $\hat{\beta} \mbox{   (}95\%\, \mbox{CI)}$ & $\hat{\pi} \mbox{   (}95\%\, \mbox{CI)}$ \\
%    \hline
%  2 observed follow-ups & 1.04 (1.03,1.05) & 205.46 (203.13,207.78) & 0.21 (0.21,0.21) \\
%  3 observed follow-ups & 1.02 (1.02,1.03) & 204.20 (202.21,206.20) & 0.21 (0.2,0.21) \\
%  4 observed follow-ups & 1.02 (1.01,1.03) & 203.39 (201.58,205.19) & 0.2 (0.2,0.2) \\
%     \hline
%  \end{tabular}
%  \caption{Component and mixing parameter estimates and confidence intervals for variable follow-up under true parameters of $\alpha=1$ and $\beta=200$ and $\pi=0.2$ based on 100 iterations of each scenario.}
% \label{tab:estVari}
% \end{table}

 % latex table generated in R 4.2.1 by xtable 1.8-4 package
 % Sat Aug 10 23:17:02 2024
 \begin{table}[ht]
 \centering
 \begin{tabular}{clll}
   \hline
   % \multicolumn{1}{c}{Scenario:}  & & & \\
  %  \hline
  % $\alpha$ & $\beta$ & $\pi$& $\hat{\alpha}\mbox{ with }95\%\, \mbox{CI}\,\, (\alpha=1\mbox{ or }2)$ & $\hat{\beta} \mbox{ with }95\%\, \mbox{CI}\,\,\, (\beta=100\mbox{ or }200)$ & $\hat{\pi} \mbox{ with }95\%\, \mbox{CI}\,\,\, (\pi=0.2)$ \\
% \hline
   Scenario & $\hat{\alpha}\mbox{   (}95\%\, \mbox{CI)}$ & $\hat{\beta} \mbox{   (}95\%\, \mbox{CI)}$ & $\hat{\pi} \mbox{   (}95\%\, \mbox{CI)}$ \\
   \hline
  {\bf Variable follow-up} &  &  &  \\
 2 observed follow-ups & 1.04 (1.03,1.05) & 205.46 (203.13,207.78) & 0.21 (0.21,0.21) \\
 3 observed follow-ups & 1.02 (1.02,1.03) & 204.20 (202.21,206.20) & 0.21 (0.2,0.21) \\
 4 observed follow-ups & 1.02 (1.01,1.03) & 203.39 (201.58,205.19) & 0.2 (0.2,0.2) \\
   &  &  &  \\
  % Variable missingness 
  % 
   \multicolumn{1}{c}{{\bf Variable missingness}} &  &    &  \\
   % \hline
 7\% missing & 1.03 (1.02,1.03) & 204.38 (202.13,206.68) & 0.21 (0.21,0.21) \\ 
   14\% missing & 1.03 (1.02,1.04) & 204.39 (202.29,206.63) & 0.21 (0.21,0.21) \\ 
   21\% missing & 1.03 (1.02,1.04) & 204.68 (202.45,206.95) & 0.21 (0.21,0.21) \\ 
    \hline
 \end{tabular}
 \caption{Component and mixing parameter estimates and confidence intervals for a variable number of observed follow-ups and variable missingnes under true parameters of $\alpha=1$ and $\beta=200$ and $\pi=0.2$ based on 100 iterations of each scenario.  Missingness simulations assume 3 observed follow-ups and the labeled average missingness among the second two follow-up periods.}
\label{tab:estVari}
\end{table}

%  7\% across 2 follow-ups & 1.04 (1.03,1.05) & 205.46 (203.13,207.78) & 0.21 (0.21,0.21) \\
%   14\% across 2 follow-ups & 1.02 (1.02,1.03) & 204.20 (202.21,206.20) & 0.21 (0.2,0.21) \\
%   21\% across 2 follow-ups & 1.02 (1.01,1.03) & 203.39 (201.58,205.19) & 0.2 (0.2,0.2) \\
% \begin{table}[ht]
% \centering
% \begin{tabular}{rlll}
%   \hline
%  & 1 & 2 & 3 \\ 
%   \hline
% 1 & 1.03 (1.02,1.03) & 204.38 (202.13,206.68) & 0.21 (0.21,0.21) \\ 
%   2 & 1.03 (1.02,1.04) & 204.39 (202.29,206.63) & 0.21 (0.21,0.21) \\ 
%   3 & 1.03 (1.02,1.04) & 204.68 (202.45,206.95) & 0.21 (0.21,0.21) \\ 
%    \hline
% \end{tabular}
% \end{table}

% $\hat{\alpha}=1, \hat{\beta}=100, \hat{\pi}=0.2$

 % % latex table generated in R 4.2.1 by xtable 1.8-4 package
 % % Fri Aug  9 23:55:54 2024
 % \begin{table}[ht]
 % \centering
 % \begin{tabular}{rlll}
 %   \hline
 %  & 2 times & 3 times & 4 times \\
 %   \hline
 % 1 &  (0.69,0.70) &  (0.90,0.91) &  (0.97,0.97) \\
 %   2 &  (0.99,0.99) &  (1.00,1.00) &  (1.00,1.00) \\
 %   3 &  (0.64,0.65) &  (0.87,0.89) &  (0.96,0.96) \\
 %   4 &  (1.00,1.00) &  (1.00,1.00) &  (1.00,1.00) \\
 %    \hline
 % \end{tabular}
 % \end{table}

 % latex table generated in R 4.2.1 by xtable 1.8-4 package
 % Wed Sep  4 23:44:05 2024
 \begin{table}[ht]
 \centering
 \begin{tabular}{rllll}
   \hline
   & \multicolumn{2}{c}{Sensitivity}& \multicolumn{2}{c}{1-Specificity} \\
Membership Threshold: &\hspace{0.5cm} 0.75 & \hspace{0.5cm} 0.95 & \hspace{0.5cm} 0.75 & \hspace{0.5cm} 0.95 \\
 % \hline 
 % 2 observed times & 0.70 (0.69,0.70) & 0.99 (0.99,0.99) & 0.64 (0.64,0.65) & 1.00 (1.00,1.00) \\
 % 3 observed times & 0.90 (0.90,0.91) & 1.00 (1.00,1.00) & 0.88 (0.87,0.89) & 1.00 (1.00,1.00) \\
 %  4 observed times & 0.97 (0.97,0.97) & 1.00 (1.00,1.00) & 0.96 (0.96,0.96) & 1.00 (1.00,1.00) \\
 %    \hline
   \hline
 2 observed follow-ups & 0.70 (0.69,0.70) & 0.64 (0.64,0.65) & 0.99 (0.99,0.99) & 1.00 (1.00,1.00) \\
   3 observed follow-ups & 0.90 (0.90,0.91) & 0.88 (0.87,0.89) & 1.00 (1.00,1.00) & 1.00 (1.00,1.00) \\
   4 observed follow-ups & 0.97 (0.97,0.97) & 0.96 (0.96,0.96) & 1.00 (1.00,1.00) & 1.00 (1.00,1.00) \\
    \hline \\ 
 \end{tabular}
 \caption{Sensitivity and specificity rates as a function of component membership threshold and number of observed follow-up periods.}
\label{tab:sens}
 \end{table}

% THINK REMOVE THIS:
% \begin{table}[ht]
% \centering
% \begin{tabular}{rrrrr}
%   \hline
%    & \multicolumn{2}{c}{Sensitivity}& \multicolumn{2}{c}{1-Specificity} \\
% Membership Threshold: & 0.8 & 0.95 & 0.8 & 0.95 \\
%  \hline 
%  2 observed follow-ups & 0.62 &   0.45 &   0.76 &   0.92 \\
%  3 observed follow-ups  & 0.73 &   0.56  &   0.81 &   0.95 \\
%  4 observed follow-ups & 0.81 &   0.72  &   0.92 &   0.98 \\
%    \hline
% \end{tabular}
% \caption{Sensitivity and specifity of dynamic clones observation under $\alpha=0.8$ and $\beta = 700$ for different membership probability thresholds and numbers of observed follow up times.}
% \label{tab:sens}
% \end{table}

% AND THINK REMOVE THIS TOO:
% \begin{table}[ht]
% \centering
% \begin{tabular}{rrrr}
%   \hline
%     \multicolumn{2}{c}{True Parameter}& \multicolumn{2}{c}{Model estimated (95\% CI)} \\
%  $\alpha$ & $\beta$ & $\hat{\alpha}$ & $\hat{\beta}$ \\
%  \hline 
%  2 &   100 &   2.1  &   97 \\
%  4 &   100  &   4.3  &   98 \\
%  2&   200  &   2.0 &   194 \\
%  4&   200  &   4.2 &   198 \\
%    \hline
% \end{tabular}
% \caption{Based on 50 iterations of each parameter pairing}
% \label{tab:est}
% \end{table}

% \begin{tabular}{|c|cc|}
% \hline
% \multirow{4}{*}{Foo} & 1 & 2 \\
% & 1 & 2 \\
% \cline{2-3}
% & 1 & 2 \\
% & 1 & 2 \\
% \hline
% \multirow{4}{*}{Bar} & 1 & 2 \\
% & 1 & 2 \\
% \cline{2-3}
% & 1 & 2 \\
% & 1 & 2 \\
% \hline
% \end{tabular}

\section{Results}
\label{sec:results}

We fit the model to a longitudinal cohort of 108 prostate cancer patients to whom androgen-deprivation therapy was applied and metastasis-directed radiation therapy (MDT) was randomized \parencite{tang_addition_2023,sherry2023peripheral}. % ludmir2024addition}.  
We performed analyses of the baseline and first follow-up cross sections, as well as baseline and two follow-up cross sections, the latter with approximately 15\% missingness across the second two follow-up periods.  The total number of productive clonotypes considered across person-times was 4.9 million.  After filtering clones to have at least 8 template reads in total across the baseline and follow-ups considered, we arrived at 62,662 clones on 97 observations for the baseline-followup analysis, and 86,381 on 104 observations for the baseline with two follow-ups analysis.  For this latter analysis, one follow-up occurred at the end of radiation for the MDT stratum and at progression for the non-MDT stratum among patients who experienced disease progression.  While only a subset of the non-MDT stratum experienced progression, one might expect it to increase model-defined dynamism because of immune response, thus pulling the non-MDT stratum toward MDT and therefore the null hypothesis with respect to that metric.  For each person-time, the total number of template reads $O_{jk}$ was calculated for use as the logged offset, giving component membership that is a function of clone variability on the scale of clonotype proportion.
% , rather than the count of template reads.  
We fit the model to these data across MDT treatment strata, and also within treatment strata to examine heterogeneity in estimated $\hat{\alpha}$, $\hat{\beta}$, and $\hat{\pi}$. Once the convergence criterion was reached, we examined cluster membership and parameter estimates.  Using a threshold component membership probability of $>0.75$ to determine the number of unique dynamic clones, we counted the number within patient and associated the measure with treatment stratum.  For the baseline-followup analysis, we additionally partitioned the number of dynamic clones into those expanding and contracting based on the trajectory of the proportion change over baseline and follow-up.  

\begin{figure}[H]
\begin{subfigure}[h]{0.48\linewidth}
    \includegraphics[width=1.01\textwidth]{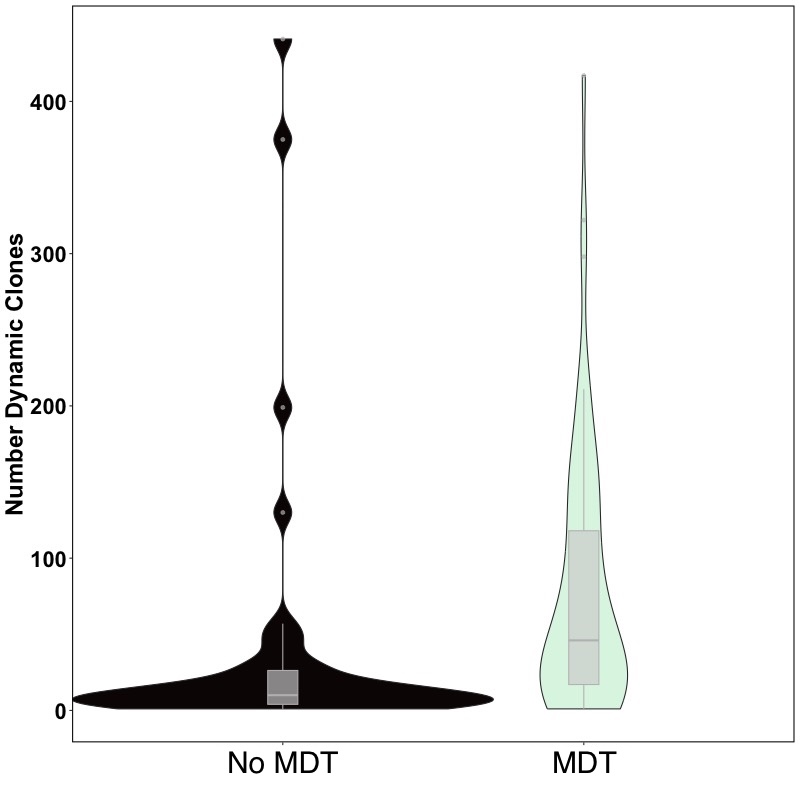} % viz_eff0_7.pdf}
\caption{Densities of the number of dynamic clones with boxplot overlay, stratified by metastasis-directed radiation therapy.}
% \label{fig:sim1}  %70}
\end{subfigure}
\hfill
\begin{subfigure}[h]{0.48\linewidth}
\includegraphics[width=1.01\textwidth]{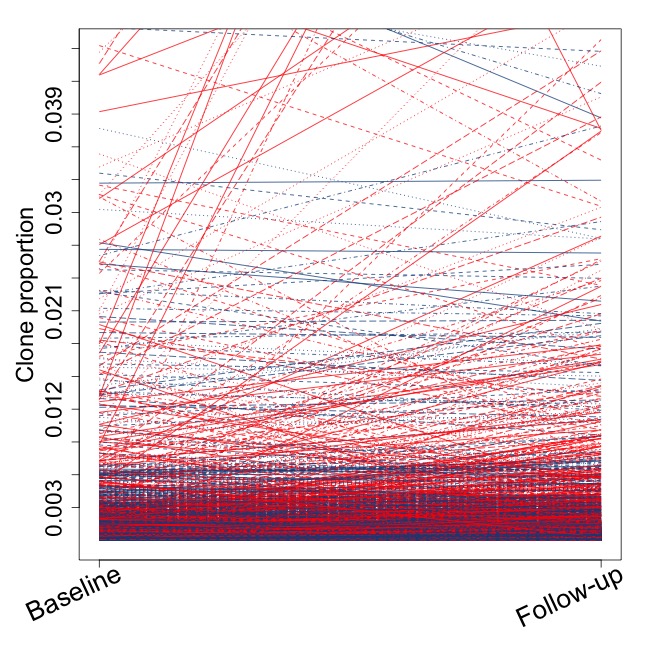} % viz_eff0_7.pdf}
\caption{Proportion clonotype over time with dynamic clones shown in red and static clones in dark blue as estimated by the model.  Static clones are downsampled by a factor of 2 for visibility of dynamic clones.}
% \label{fig:sim1}  %70}
\end{subfigure}
\caption{Two follow-up period analysis of the baseline and first follow-up cross sections using variance component mixture model. }
% A log-linear model of the number of dynamic clones regressed on radiation stratum yields a parameter estimate of 0.67, suggesting an expected near doubling of the number of dynamic clones for those in the radiation stratum versus not ($p<10^{-8}$). $\chi^2$ tests based on dichotomizing whether observations have more or fewer than 50 dynamic clones likewise shows a significant association with radiation therapy stratum ($p<0.001$).}
\label{fig:analysis}
\end{figure}

\begin{figure}[H]
\begin{subfigure}[h]{0.48\linewidth}
\includegraphics[width=1.01\textwidth]{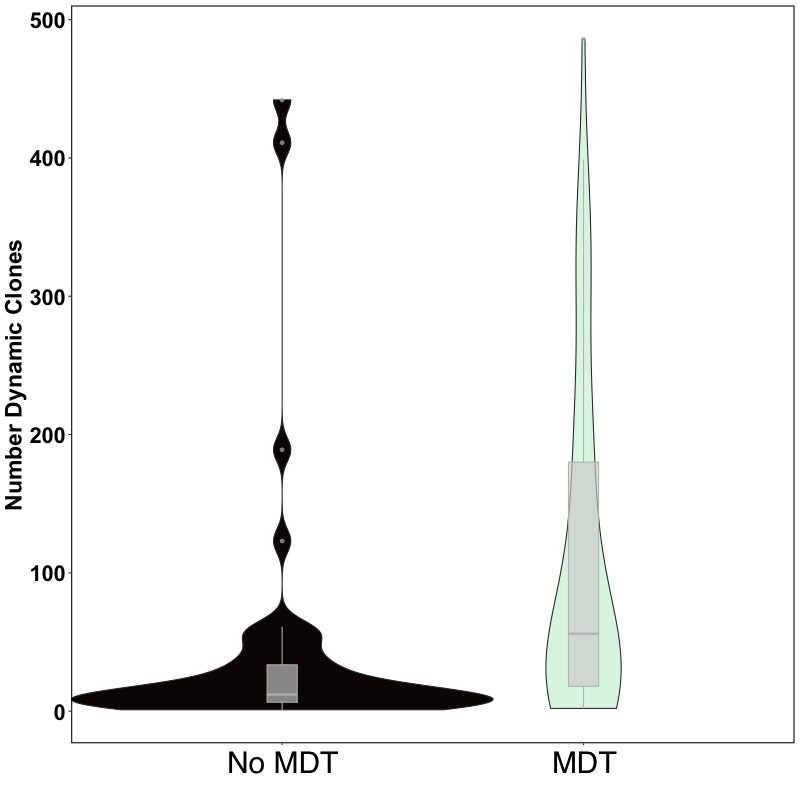} % viz_eff0_7.pdf}
\caption{Densities of the number of dynamic clones with boxplot overlay, stratified by metastasis-directed radiation therapy.}
% \label{fig:sim1}  %70}
\end{subfigure}
\hfill
\begin{subfigure}[h]{0.48\linewidth}
\includegraphics[width=1.01\textwidth]{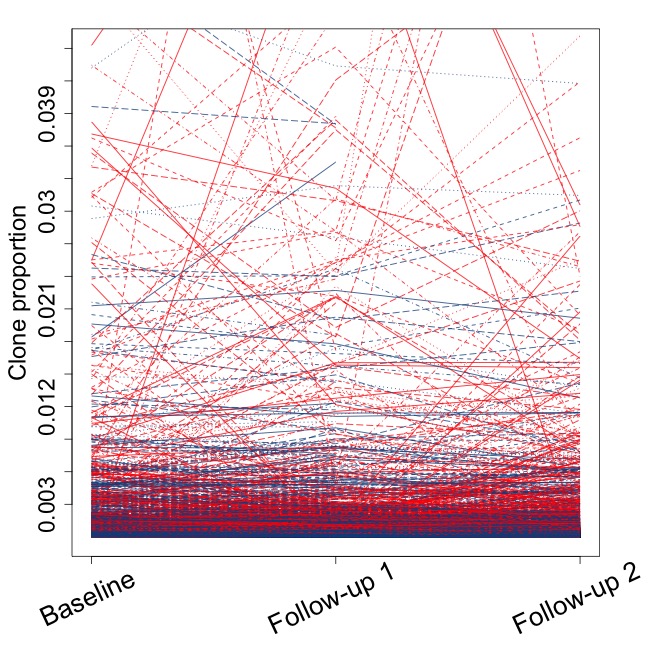} % viz_eff0_7.pdf}
\caption{Proportion clonotype over time with dynamic clones shown in red and static clones in dark blue as estimated by the model.  Static clones are downsampled by a factor of 2 for visibility of dynamic clones.}
% \label{fig:sim1}  %70}
\end{subfigure}
\caption{Three follow-up period analysis using variance component mixture model.}
% A log-linear model of the number of dynamic clones regressed on radiation stratum yields a parameter estimate of 0.67, suggesting an expected near doubling of the number of dynamic clones for those in the radiation stratum versus not ($p<10^{-8}$). $\chi^2$ tests based on dichotomizing whether observations have more or fewer than 50 dynamic clones likewise shows a significant association with radiation therapy stratum ($p<0.003$).}
\label{fig:analysis2}
\end{figure}

\begin{figure}[H]
\begin{subfigure}[h]{0.46\linewidth}
\includegraphics[width=1.01\textwidth]{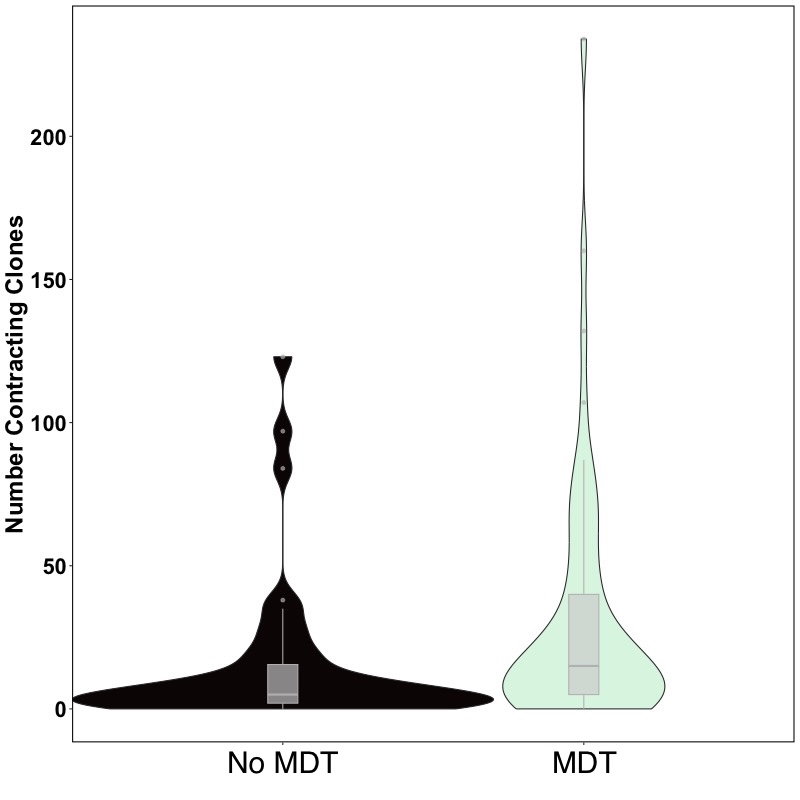} % viz_eff0_7.pdf}
\caption{Contracting clone count densities with boxplot overlay, stratified by metastasis-directed radiation therapy.}
% \label{fig:sim1}  %70}
\end{subfigure}
\hfill
\begin{subfigure}[h]{0.46\linewidth}
\includegraphics[width=1.01\textwidth]{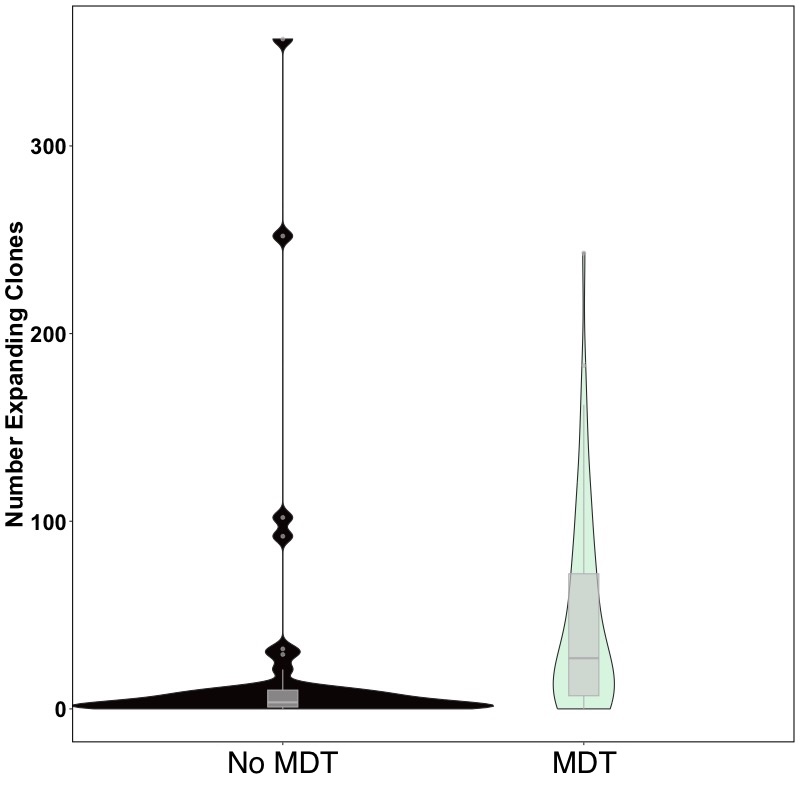} % viz_eff0_7.pdf}
\caption{Expanding clone count densities with boxplot overlay, stratified by metastasis-directed radiation therapy.}
% \label{fig:sim1}  %70}
\end{subfigure}
\caption{Analysis of the baseline and follow-up cross sections using variance component mixture model for number of contractions and number of expansions.}
% A log-linear model of the number of dynamic clones regressed on radiation stratum yields a parameter estimate of ....}
\label{fig:analysis3}
\end{figure}

Results indicate a significant increase in the number of dynamic clones under MDT for both longitudinal analyses. Figures \ref{fig:analysis}a), \ref{fig:analysis2}a), and \ref{fig:analysis3} all show qualitative marked increases in the number of dynamic, expanding, and contracting clones in the MDT treatment stratum, which are formally tested and found significant in nearly all cases by $\chi^2$ tests and log-linear models (Table \ref{tab:threshold}).  The non-MDT stratum had a small number of outlier observations that exhibited a large number of dynamic clones, likely pulling test statistics more toward the null hypothesis of no treatment strata differences than would have otherwise been observed.  

One observes in Figures \ref{fig:analysis}b) and \ref{fig:analysis2}b)  that dynamic clones tend to exhibit significant slopes as the clonotype proportion changes dramatically across longitudinal follow-up, in contrast to static clones which are flatter.  One also sees that the higher the average clonotype proportion, the larger absolute change in proportion is needed for a clone to be considered dynamic, incorporating expected biological variability into the model in assessing dynamic behavior.  

Within MDT treatment strata, models reveal some heterogeneity in estimated $\alpha$ and $\beta$ as shown in Table \ref{tab:paramest}, suggesting differences in shape of clonal dynamics over time as a function of treatment.  Since $\alpha$ and $\beta$ govern both the dynamic and static mixture components, interpretation is not straightforward, though the implication of estimates is that among the MDT stratum, clonotype proportions tend to be smaller and have smaller variance than the non-MDT stratum across time and components.  Since clonotype proportions are sensitive to the number of unique clones within observation-times because they must sum to one, interpretation might indicate a tendency toward less ``focus'' in dynamism among MDT stratum clones as those useful in immune response expand and those that are not contract.

\begin{table}[ht]
\centering
\begin{tabular}{lllll}
  \hline
  &  \multicolumn{3}{c}{2 observed follow-ups (FU's)} & \multicolumn{1}{c}{3 FU's} \\
  \multicolumn{1}{r}{Clone group:} &  Dynamic & Expansions & Contractions & Dynamic \\
 \hline 
 $\chi^2$ test p-value &  0.0007  &  0.0003  &   0.13  &   0.001 \\
  log-linear parameter (p-value) & 0.68 ($10^{-8}$)  &  0.66 ($10^{-8}$) &   0.71 ($10^{-8}$)  &   0.98 ($10^{-8}$) \\
   \hline
\end{tabular}
\caption{Hypothesis tests for dynamic clone stratification by MDT treatment stratum.  $\chi^2$ tests are based on dichotomizing at 50 total dynamic clones or 25 total expanding/contracting clones depending on the case.  The log-linear model parameter estimate indicates the log expected proportion increase in the number of dynamic, expanding, or contracting clones, depending on the case, for the MDT stratum as compared to no MDT.}
\label{tab:threshold}
\end{table}

 % latex table generated in R 4.2.1 by xtable 1.8-4 package
 % Fri Aug  9 23:54:56 2024
 \begin{table}[ht]
 \centering
 % \begin{tabular}{ccccccc}
 \begin{tabular}{llll|lll}
   \hline
% \multicolumn{2}{c}{Sensitivity}
  & \multicolumn{3}{c}{2 observed follow-ups} & \multicolumn{3}{c}{3 observed follow-ups} \\
   % \hline
  & Combined & No MDT & MDT & Combined& No MDT & MDT  \\
   \hline
 $\hat{\alpha}$ & 0.35 & 0.36 & 0.34  & 0.33 & 0.37 & 0.32  \\
  $ \hat{\beta}$& 640.00 & 603.41 & 658.44  & 715.33 & 648.81 & 752.05 \\
  $ \hat{\pi}$& 0.23 & 0.22 & 0.24  & 0.20 & 0.19 & 0.20  \\
    \hline
 \end{tabular}
 \caption{Model parameter estimates for two and three study follow-ups, across and within the metastasis-directed radiation therapy (MDT) and lack of MDT (No MDT) treatment strata.}
\label{tab:paramest}
 \end{table}

\vspace{0.5cm}

\section{Discussion}
\label{sec:discussion}

In this work we placed T-cell receptor clonal dynamics in a mixture model framework grounded in Bayesian hierarchy and argued the approach can be a useful tool for understanding these high-dimensional data that duly acknowledges expected biological variability while incorporating fundamental model features like variable longitudinal follow-up and missingness.  Since receptors shared across subjects are rare relative to those unique to each subject, traditional statistical high-dimensional regression tools are not as useful and one must try to relate summarizations of clonal population dynamics to subject phenotype, in our case a count of model-determined dynamic clones.  Additionally, partitioning dynamic clones into those that expand and contract as a function of their trajectory further allows one to examine the kind of dynamism most sensitive to intervention, here metastasis-directed radiation therapy.  Early identification of clinically relevant dynamic clones shortly following interventions like metastasis-directed radiation therapy may facilitate treatment decision-making and subsequent therapeutic approaches, representing a promising actionable biomarker underlying host immunobiology. In addition, resolution of the systemic TCR repertoire through this approach is more practical and feasible than serial tissue biopsies, particularly as co-culturing assays of T cells with tumor tissue are laborious, resource intensive, and can be challenging to scale.  
% Statistical development at this interface to better understand clonal dynamics is relevant to patient decision making in this setting.  

The mixture of negative multinomial and product of negative binomial components arrived at through Gamma-Poisson conjugacy was both theoretically convenient and reflective of TCR clonal dynamics and therefore a promising means of modeling these data.  However, future work may well attempt more flexible error model alternatives for which computationally intensive fitting procedures are necessary.  Since the single parameter Poisson distribution equates the mean and variance, enforcing the relationship between clone frequency and degree of variability constituting ``dynamism'', a model allowing one to specify that relationship could help clinicians and subject matter experts place greater weight on, say, dynamism among the low-frequency group of clones or vice versa, depending on hypothesized biological mechanism.  The mean-variance equivalence of the Poisson distribution does have advantages in that since coverage increases in the root of the variance, one is better powered to identify dynamism among higher frequency clones for less variability relative to that mean, consistent with a biological prior that higher frequency clones are inherently more interesting and should be easier to label as dynamic.
% for a fixed change in proportion.  
Still, exploring sensitivity to different and more flexible mean-variance specifications could improve characterization of the dynamism associated with subject phenotype.  

\section{Acknowledgments}

This work was supported by CPRIT grant Rp180140, the Andrew Sabin Family Fellowship, and the National Cancer Institute grant P30 CA016672.

\printbibliography

% \printbibliography 
% \bibliographystyle{abbrvnat} 
% % \bibliography{refs_fdr_splines7_29}
% \bibliography{tcr7_11.bib}

\pagebreak

\end{document}